\documentstyle[12pt]{article}
\def\thalf{{\textstyle{\frac{1}{2}}}}
\def\tquar{{\textstyle{\frac{1}{4}}}}
\def\pj{\hspace{-.26cm}}
\def\Tr{{\rm Tr}}
\def\s{\sigma}
\def\po{\varepsilon}
\def\beq{\begin{equation}}
\def\eeq{\end{equation}}
\def\gord{$ \raisebox{-.3ex}{$\stackrel{>}{_{\sim}}$} $}

\begin{document}
\begin{titlepage}
\begin{center}
{\Large \bf Theoretical Physics Institute \\
University of Minnesota \\}  \end{center}
\vspace{0.15in}
\begin{flushright}
TPI-MINN-92/37-T
\end{flushright}
\vspace{0.2in}
\begin{center}
{\Large \bf Temperature Dependence of Electric and Magnetic Gluon
Condensates \\}
\vspace{0.4in}
{\bf V.L. Eletsky$^{\dagger}$,} \\
Theoretical Physics Institute, University of Minnesota  \\
Minneapolis, MN 55455 \\
\vspace{0.2in}
{\bf P.J. Ellis and J.I. Kapusta \\}
School of Physics and Astronomy, University of Minnesota \\
Minneapolis, MN 55455 \\
\vspace{0.2in}
{\bf   Abstract  \\ }
\end{center}

The contribution of Lorentz non-scalar operators to finite temperature
correlation
functions is discussed. Using the local duality approach for the one-pion
matrix element of a product of two vector currents, the temperature dependence
of the average gluonic stress tensor is estimated in the chiral limit to be
$\langle{\bf E}^2 +{\bf B}^2\rangle_{T}=\frac{\pi^2}{10}bT^4$.
At a normalization point $\mu=0.5$ GeV we obtain $b\approx 1.1$.  Together with
the known temperature dependence of the Lorentz scalar gluon condensate we are
able to infer $\langle{\bf E}^2\rangle_T$ and $\langle{\bf B}^2\rangle_T$
separately in the low-temperature hadronic phase.

\vskip1in
\hrule height .2pt width 3in
\vskip1pt
\noindent$^{\dagger}$Permanent address: Institute of Theoretical and
Experimental Physics, Moscow 117259, Russia.
\end{titlepage}

	Correlators of currents with the quantum numbers of hadrons are known to
be useful to obtain information about the masses and couplings of hadrons; they
are employed in the QCD sum rule approach and in lattice calculations. In both
approaches the correlators are considered at large Euclidean distances or
imaginary times where the dominant contribution comes from the lowest
state with the corresponding quantum numbers. QCD sum
rules give predictions also for form factors and structure
functions of hadrons. (For a recent review  of
applications of  QCD correlation functions see ref. \cite{shu}.)

In recent years there has been increasing interest in finite temperature QCD
and hadronic physics  due to the expectation that at high enough temperatures
the QCD vacuum, specified by nonperturbative condensates of quark and gluon
fields, will ``melt" and undergo a transition to a quark-gluon plasma.
Melting is
usually understood in the sense that chiral symmetry restoration and
deconfinement take place. The former means that with increasing temperature
quark condensates evaporate, while the latter means that hadrons do not
represent stable degrees of freedom.
It was shown by Leutwyler and his collaborators \cite{leut}
using the chiral Lagrangian approach
that the quark condensate indeed decreases with rising
temperature.
{}From the usual QCD sum rules at $T=0$
it is well known that the properties of hadrons are, to a large
extent, determined by  nonperturbative quark and gluon condensates  \cite{svz}.
Naturally, a large number of papers were devoted to the generalization of
QCD sum rules to finite temperature in attempts to
relate the temperature  dependence of the hadronic spectrum to the temperature
dependence of the condensates (see, e.g. \cite{bs,dei,ahz}). In this case the
vacuum average of the product of currents becomes the Gibbs
average over the thermal ensemble. To calculate the Gibbs average
one must choose a basis for the states.
As argued in refs. \cite{dei,els} at
temperatures which are much less than the energy scale of
confinement the appropriate basis is that of hadronic states, rather than the
quark-gluon basis used in early papers on the subject
(see, e.g. ref. \cite{bs}).
Using this basis it was also shown \cite{dei} that at low $T$ the
thermal correlators are expressed
as a mixture of zero-temperature correlators with different parity.
It is also clear that if the operator product expansion (OPE) is applied
to a thermal correlator
then the temperature dependence appears only in the
matrix elements of the operators (condensates), the coefficient functions
being obtained through a perturbative calculation at $T=0$. QCD sum rules at
low temperature were recently reexamined along these lines in
ref. \cite{hkl}.\footnote{We thank T. Hatsuda for drawing our attention to
this paper.} At high temperatures, corresponding
to the quark-gluon plasma, the calculation of thermal correlators should be
performed in a basis consisting of quark and gluon states.
In this case the perturbative temperature-dependent parts of
the condensates due to quarks and gluons from the thermal ensemble
may be included in the coefficient functions \cite{ahz}.

Thus the QCD sum rule method, understood as a tool to get
information about the imaginary parts of correlators via analyticity,
seems to be tractable
both at very low and very high temperatures, but not in  the region of a phase
transition where a drastic  rearrangement of the spectrum takes place.

An additional feature of finite temperature sum rules is the
appearence of new condensates due to Lorentz non-scalar operators; these were,
of course, present in the OPE, but gave zero contribution when averaged over
the
vacuum. At finite temperatures Lorentz invariance is broken and
these operators should contribute \cite{shu,ahz}. The same applies to the case
of finite density \cite{fd}. However, each of these new condensates is
an unknown nonperturbative parameter. In principle they may be fixed
from the physical spectral densities of the correlators, just as in the
zero temperature case the now well-established condensates were fixed by the
hadronic spectrum.

Consider the correlator of two isovector vector currents at
finite temperature $T$ and euclidean momentum $q$, where $T^2\ll Q^2=-q^2$ and
$Q^2\gord1$ GeV$^2$:
\beq
i\int d^4 x e^{iqx} \sum_{n}\langle n|{\cal T}j_{\mu}(x)j_{\nu}(0)
e^{(\Omega-H)/T}|n\rangle=
(g_{\mu\nu}q^2-q_{\mu}q_{\nu})C_1(q,T)+u^{t}_{\mu}u^{t}_{\nu}C_2(q,T)\;,
\label{def}
\eeq
where ${\cal T}$ denotes a time-ordered product,
$j_{\mu}=\thalf(\bar{u}\gamma_{\mu}u-\bar{d}\gamma_{\mu}d)$,
$u^{t}_{\mu}=u_{\mu}-(u\cdot q)q_{\mu}/q^2$ is the transverse part
of the heat bath four-velocity $u_{\mu}$ and
$\Omega=-T\log (\sum_{n}\langle n|e^{-H/T}|n\rangle)$.
Eq. (\ref{def}) is the most general expression compatible with conservation
of the vector current. The Lorentz invariance breaking term
proportional to $u^{t}_{\mu}u^{t}_{\nu}$ must be absent at $T=0$. This means
that $C_2(q,T)$ goes to zero as $T\to 0$, while $C_1(q,T)$ becomes
the usual zero temperature correlator.
Notice that eq. (\ref{def}) may be considered to be the amplitude for forward
scattering of a virtual photon by the heat bath. Then the imaginary parts of
$C_1$ and $C_2$ are the structure functions of deep inelastic scattering of
leptons by the heat bath ($u^{T}_{\mu}$ is similar to the transverse component
of the target momentum, $p_{\mu}-(p\cdot q)q_{\mu}/q^2$).

At low $T$, when the contributions
from all particles except pions  are exponentially
suppressed in the Gibbs average, the functions $C_1$ and $C_2$ may be
estimated by expanding in the
density of thermal pions. In the first order of this expansion only matrix
elements over one-pion states are taken into account. This approximation
was made in ref. \cite{dei} for $C_1$.
The one-pion matrix elements were estimated via PCAC
and current algebra. It was shown that $C_1$ and its counterpart from the
axial channel are given
by $T$-dependent mixtures of their zero temperature values
and, as a result, the corresponding screening lengths tend to
converge with increasing temperature.

The purpose of the present letter is to estimate $C_2$. Let us  start from the
one-pion matrix element in the chiral limit
\beq
i\int d^4 x e^{iqx}\langle\pi (p)|{\cal T}j_{\mu}(x)j_{\nu}(0)|\pi
(p)\rangle\;,
\label{me}
\eeq
where we assume $p\sim T\ll Q$, since eq. (\ref{me}) is to be integrated over
$p$ with Bose occupation probabilities.
If $\hat O_{\mu_{1}\mu_{2}...\mu_{n}}$ is an operator of Lorentz spin $n$, then
the matrix element
$\langle\pi (p)|\hat O_{\mu_{1}\mu_{2}\ldots\mu_{n}}|\pi (p)\rangle\propto
p_{\mu_{1}}p_{\mu_{2}}
\ldots p_{\mu_{n}}$, and cannot be reduced via PCAC to a vacuum matrix
element. It is clear that at low temperatures, $T\ll Q$,
the main contribution to $C_2$ comes from operators of lowest spin, namely
spin 2.
In leading twist there are two spin 2 operators which are related to the
energy-momentum tensor:
\begin{eqnarray}
\theta^q_{\mu_1\mu_2}&\pj=&\pj \thalf i(\bar q\gamma_{\mu_1}D_{\mu_2}q+
\bar q\gamma_{\mu_2}D_{\mu_1}q)\quad,\quad  q=u,d,s\ldots\nonumber\\
\theta^G_{\mu_1\mu_2}&\pj=&\pj G_{\mu_1\alpha}^{a}
G_{\phantom{a\alpha}\mu_2}^{a\alpha}-
\tquar g_{\mu_1\mu_2}G_{\beta\alpha}^{a}G^{a\alpha\beta}\;,
\label{thetaG}
\end{eqnarray}
where $D_{\mu}$ is the covariant derivative.
Graphs which correspond to the contributions of these operators to the matrix
element in eq. (\ref{me}) are shown in fig. 1.
If the normalization point for the operators is taken to be $\mu^2=Q^2$, then
the operator $\theta^G_{\mu_1\mu_2}$ does not contribute to the OPE in the
leading $\log$ approximation, and the contribution
of twist 2, spin 2 operators to eq. (\ref{me}) involves
\beq
\frac{1}{Q^2}\langle\pi (p)|\theta^u_{\mu\nu}+\theta^d_{\mu\nu}|\pi (p)\rangle=
\frac{1}{Q^2}\langle\pi (p)|\theta^{\rm tot}_{\mu\nu}-\theta^G_{\mu\nu}|\pi
(p)\rangle\;.
\label{tw2}
\eeq
Here we neglected the contributions of heavy quarks. The matrix element of the
total energy-momentum tensor is
$\langle\pi (p)|\theta^{\rm tot}_{\mu\nu}|\pi (p)\rangle=2p_{\mu}p_{\nu}$
(the states are normalized such that
$\langle\pi (p)|\pi (p')\rangle=(2\pi)^{3}2E\delta^{(3)}({\bf p}-{\bf p'})$),
while the matrix element of the gluon energy-momentum tensor
\beq
\langle\pi (p)|\theta^G_{\mu\nu}|\pi (p')\rangle=bp_{\mu}p_{\nu}
\label{defb}
\eeq
contains an unknown constant $b$. This constant is related to the
matrix element of the energy density of the gluon field
\beq
b=\frac{1}{2{\bf p}^2}\langle\pi (p)|{\bf E}^2+
{\bf B}^2|\pi (p)\rangle_{\mu =Q}\;.
\label{b}
\eeq
Note that $b$ depends on the normalization point, $\mu$, in the operator
product
expansion. This dependence will be discussed later.

Let us try to estimate $b$ within a quark-hadron duality approach,
saturating the amplitude of eq. (\ref{me}) by hadrons,
$\langle\pi|{\cal T}j_{\mu}(x)j_{\nu}(0)|\pi\rangle=
\sum_{n}\langle\pi|j_{\mu}(x)|n
\rangle\langle n|j_{\nu}(0)|\pi\rangle$.
Focussing on spin 2 contributions to $C_2$, we then have
\beq
\frac{2-b}{Q^2}+\frac{c}{Q^4}+\ldots=
\frac{1}{\pi}\int_{0}^{\infty} ds\frac{\rho (s)F^{2}_{n}(Q^2)}{s+Q^2}\;,
\label{disp}
\eeq
where $F_{n}(Q^2)$ is the part of the form factor
$\langle\pi(p)|j_{\mu}|n(p+q)\rangle$ proportional to $p_{\mu}$
and $\rho(s)$ is the spectral
density in the $s$-channel. The states $|n\rangle$ are normalized as in
eq. (\ref{defb}), the $n$-state contribution to $\rho (s)$ being
$\pi\delta(s-m_{n}^2)$. On the l.h.s. of eq. (\ref{disp}) the term
$c/Q^4$ denotes the contribution of three different spin 2, twist 4
operators \cite{sv} whose individual
contributions cannot be separated. The constants $b$ and $c$ are considered as
parameters to be
fitted. The ellipsis in eq. (\ref{disp}) corresponds to spin 2 terms of higher
twist. Note that eq. (\ref{disp}) is just the sum rule for the second moment
of the deep inelastic structure function, $\int_{0}^{1}F_{2}(x,Q^2)dx$,
divided by $Q^2$. It is valid in the asymptotic region, $Q^2\to\infty$, with
all
higher states in the $s$-channel equally important in this region.
Our goal here is to see whether eq. (\ref{disp}) can be satisfied in a region
of intermediate $Q^2\sim 1$ GeV$^2$ where the r.h.s. may be approximated by the
contribution of a few low-lying
states\footnote{In ref. \cite{giko}
the transverse photon structure function in
the region of intermediate $x$ was calculated starting from the
$VVVV$ four-point correlation function, using the OPE in the photon
virtuality $p^2$ and extrapolating to $p^2=0$. One could think of doing the
same thing for the pion structure function, starting from the $AVVA$
correlator.
It can be shown, however, that just as in the case of the longitudinal photon
structure function, there are difficulties in the extrapolation to
{\it on-shell} pions.
The $AVVA$ box diagram also cannot be used, via a triple dispersion relation,
to model the continuum contribution to the real part of the forward
scattering amplitude in the usual manner because of the zero momentum
transfer in the $t$-channel.}.

First consider the case of charged pions. The lowest states in
the $s$-channel are the $\pi$ and $a_1(1260)$ mesons. Assuming $\rho$-dominance
for
the form factors (which is known to be a good approximation for the pion
form factor up to $Q^2\simeq 2$ GeV$^2$), $\langle\pi|j_{\mu}|n\rangle=
-\frac{m^{2}_{\rho}}{g_{\rho}} \,\po^{\rho}_{\mu}\langle\pi\rho|n\rangle
(Q^2+m_{\rho}^2)^{-1}$, where
$\po^{\rho}_{\mu}$ is the $\rho$-meson polarization vector and
$g_{\rho}^2/4\pi\simeq 2.9$. We obtain for the r.h.s. of eq. (\ref{disp})
\begin{eqnarray}
\frac{8m_{\rho}^4}{(Q^2+m_{\rho}^2)^2}
\biggl[ \frac{1}{Q^2}+\frac{1}{4m_{a_1}^{2}g_{\rho}^{2}(Q^2+m_{a_1}^{2})}
\Bigl\{g_{a_1\rho\pi}^2&\pj+&\pj g_{a_1\rho\pi}h_{a_1\rho\pi}(Q^2-m_{a_1}^{2})
\nonumber\\&&\quad+\tquar h_{a_1\rho\pi}^2(Q^2+m_{a_1}^{2})^2\Bigr\} \biggr].
\label{pia}
\end{eqnarray}
Here we used
\beq
\langle\pi^{+}(p)\rho^{0}(q)|\pi^{+}(p+q)\rangle=
g_{\rho\pi\pi}\;\po^{\rho\ast}\cdot(2p+q)
\label{prp}
\eeq
and
\beq
i\langle\pi^{+}(p)\rho^{0}(q)|a_{1}^{+}(p+q)\rangle=
g_{a_1\rho\pi}\;\po^{\rho\ast}\cdot \po^{a_1}+
h_{a_1\rho\pi}\;\po^{\rho\ast}\cdot(p+q)\;\po^{a_1}\cdot p\;.
\label{pra}
\eeq
The notation corresponds to that of ref. \cite{hol}. Note that
$g_{\rho\pi\pi}=g_{\rho}$ within the $\rho$-dominance approach.
The couplings $g_{a_1\rho\pi}$ and $h_{a_1\rho\pi}$ cannot, of course, be
determined from the $a_1$ width alone. To this end we use an effective chiral
Lagrangian with spin 1 mesons \cite{gks,hol,mei}.
In this approach the constants in question are expressed in terms of
parameters of
this Lagrangian which are fitted to reproduce masses and widths.
The Lagrangian in question contains a massive Yang-Mills part and two higher
derivative terms
\begin{eqnarray}
{\cal L}_{AV\phi}&=&-\thalf \Tr(F_{\mu\nu }^LF^{L\mu\nu}+
F_{\mu\nu }^RF^{R\mu\nu })+
m_0^2 \Tr(A_{\mu}^LA^{L\mu}+A_{\mu}^RA^{R\mu}) \nonumber  \\
& &\mbox{}-i\xi \Tr(D_{\mu}UD_{\nu}U^{\dagger}F^{L\mu\nu}+
D_{\mu}U^{\dagger}D_{\nu}UF^{R\mu\nu})+
\s \Tr F_{\mu\nu}^LUF^{R\mu\nu}U^{\dagger},
\label{L}
\end{eqnarray}
where $U=\exp (2i\phi/F_{\pi})$, $\phi=\phi^{a}\tau^a/\sqrt{2}$,
$A_{\mu}^L=\frac{1}{2}(V_{\mu}+A_{\mu})$,
$A_{\mu}^R=\frac{1}{2}(V_{\mu}-A_{\mu})$,
$F_{\mu\nu}^{L,R}=\partial_{\mu}A_{\nu}^{L,R}-\partial_{\nu}A_{\mu}^{L,R}-
ig\left[ A_{\mu}^{L,R},A_{\nu}^{L,R}\right]$
and the covariant derivative
$D_{\mu}U=\partial_{\mu}U -igA_{\mu}^LU+igUA_{\mu}^R$. The
quadratic piece of this Lagrangian is non-diagonal in $\partial_{\mu}\phi$ and
$A_{\mu}$. After diagonalization the physical masses are given by
\begin{equation}
m_{V}^2= m_{\rho}^2=\frac{m_{0}^2}{1-\s}\quad;\quad
m_{A}^2= m_{a_1}^2=\frac{1}{1+\s}\left(m_{0}^2+\frac{g^2 F_{\pi}^2}{4}
\right)\;,
\label{m}
\end{equation}
and $F_{\pi}$ is related to the physical coupling
$\tilde{F}_{\pi}=135$ MeV through
\beq
\tilde{F}_{\pi}=ZF_{\pi}\;, ~~~~
Z^2=1-\frac{g^2\tilde{F}_{\pi}^2}{4m_{0}^2}
=\frac{1-\s}{1+\s}\:\frac{m_{V}^2}{m_{A}^2}\;.
\label{z}
\eeq
The couplings $g_{a_1\rho\pi}$, $h_{a_1\rho\pi}$ and $g_{\rho\pi\pi}$ are
expressed through $g$, $\xi$, $\s$  and the meson masses by
\begin{eqnarray}
h_{a_1\rho\pi}&\pj=&\pj-\frac{2Z^2}{\tilde{F}_{\pi}}
\left(\frac{2}{1-\sigma^2}\right)^{\!\frac{1}{2}}(\sigma+g\xi)\;,\\
g_{a_1\rho\pi}&\pj=&\pj\thalf(m_{V}^2+m_{A}^2-m^2_{\pi})h_{a_1\rho\pi}
+\frac{m_V^2}{\tilde{F}_{\pi}}\left(\frac{2}{1-\sigma^2}\right)^{\!\frac{1}{2}}
\left[(1-\sigma)(1-Z^2)+2g\xi Z^2\right]\;,\\
g_{\rho\pi\pi}&\pj=&\pj\frac{g}{\sqrt{2(1-\sigma)}}\left[ 1-\thalf(1-Z^2)
+\frac{g\xi}{(1-\sigma)}\:\frac{Z^4}{(1-Z^2)}\right]\;.
\end{eqnarray}
Here we retain a non-zero mass for the pion for the purposes of fitting the
coupling constants.
The widths are expressed through these couplings in the following
way\footnote{We note that the {\it minus} sign in eq. (\ref{gammaa1})
is correct in contrast to  refs. \cite{hol,mei} which are
written with an incorrect {\it plus} sign.}
\beq
\Gamma_{\rho\rightarrow\pi\pi}=\frac{1}{6\pi m_{\rho}^2}|q_{\pi}|^3
g_{\rho\pi\pi}^2\;,
\label{gammarho}
\eeq
and
\beq
\Gamma_{a_{1}\rightarrow\rho\pi}=\frac{|q_{\pi}|}{12\pi m_{a_1}^2}
\left[ 2g_{a_1\rho\pi}^2+
\left( \frac{E_{\rho}}{m_{\rho}}g_{a_1\rho\pi}-
\frac{m_{a_1}}{m_{\rho}}|q_{\pi}|^2 h_{a_1\rho\pi}\right) ^{\!2}\right]\;.
\label{gammaa1}
\eeq
With the four available parameters $g$, $\s$, $\xi$ and $m_0$ it is possible
to fit both the masses and the widths of the $\rho$ and the $a_1$ \cite{mei}.
We have refitted these parameters using a recent value of the width,
$\Gamma_{a_1}=400$ MeV \cite{pdg,song}.
There are two possible solutions:
\begin{eqnarray}
(A)~~~~\sigma&\pj=&\pj 0.340,~~~~\xi= 0.446,~~~~g=8.37 \;,\nonumber \\
(B)~~~~\sigma&\pj=&\pj-0.291,~~~~\xi=0.0585,~~~~g= 7.95 \;.
\label{s2}
\end{eqnarray}
which correspond to
\begin{eqnarray}
(A)~~~~g_{a_1\rho\pi}&\pj=&\pj-5.42\ {\rm GeV},~~~h_{a_1\rho\pi}=-16.7\
{\rm GeV}^{-1},~~~~\gamma=0.52\;,\nonumber  \\
(B)~~~~g_{a_1\rho\pi}&\pj=&\pj 4.25\ {\rm GeV},~~~h_{a_1\rho\pi}=-2.05\
{\rm GeV}^{-1},~~~~\gamma=0.33\;.
\label{os}
\end{eqnarray}
Here the quantity $\gamma$ is the ratio of polar- and axial-vector
contributions to radiative pion decay. Both solutions
are reasonably consistent with the positive experimental value of $\sim0.4$
discussed by Holstein \cite{hol}. However
it can be shown that the {\it opposite-sign} solution, $(B)$, is
excluded by the QCD sum rule estimates of Ioffe and Smilga \cite{is} for
the two form factors entering
the non-diagonal matrix element $\langle a_1|j_{\mu}|\pi\rangle$.
They use couplings $g_1$ and $g_2$
to parameterize these form factors in a $\rho$-dominance approach,
and the relation to $g_{a_1\rho\pi}$
and $h_{a_1\rho\pi}$ is given by
\beq
g_{a_1\rho\pi}=g_{1}m_{a_1},~~~h_{a_1\rho\pi}=\frac{2}{m_{a_1}}
\left[ g_1+\frac{m_{\rho}^2}{m_{a_1}^2}g_2\right]\;.
\label{g1g2}
\eeq
While the absolute values of $g_1$ and $g_2$ obtained in ref. \cite{is} contain
large uncertainties, they are definitely of the same sign, thus ruling out
solution $(B)$. Therefore we choose the {\it like-sign} solution $(A)$.

We shall display our results for the r.h.s. of eq. (\ref{disp}) multiplied by
$Q^4$, which according to the l.h.s should give the
linear relation $(2-b)Q^2+c$. The results from eq. (8)
are given by the dashed line for the charged pion case in fig. 2. It is seen
that there is a good linear dependence for $ Q^2\geq 0.9$ GeV$^2$ .
We cannot use values of $Q^2$ larger than plotted in the figure since higher
states, which are not accounted for,
become important and $\rho$-dominance is not applicable either.
There is an excited pion state $\pi^{\ast} (1300)$ which
may contribute for the values of $Q^2$ in question. Its coupling to $\rho\pi$
defined through
$\langle\pi^{\ast}|\rho(q)\pi(p)\rangle=g^{\ast}\;\po^{\rho}\cdot p$ may be
roughly estimated using the rather uncertain data \cite{pdg} on the width,
$\Gamma_{\pi^{\ast}}^{\rm tot}=200-600$ MeV and $\Gamma_{\pi^{\ast}\to\pi\rho}=
\frac{1}{3}\Gamma_{\pi^{\ast}}^{\rm tot}$. This gives $g^{\ast}\approx 5$. The
contribution of the $\pi^{\ast}$ to the r.h.s. of eq. (\ref{disp}) is then
\beq
\frac{2g^{\ast 2}m_{\rho}^4}
{g_{\rho}^2 (Q^2+m_{\rho}^2)^2 (Q^2+m^2_{\pi^{\ast}})}\;.
\label{ep}
\eeq
The result of taking into account the $\pi^{\ast}$ is shown in fig. 2 by the
full line. It is clear that the effect of the $\pi^{\ast}$ is
quite small. By fitting a straight line to the curve we estimate $b=1.14$ and
$c=1.14$ GeV$^2$.

The matrix element of the gluon field energy density,
eq. (\ref{b}), must be the same for charged and neutral pions. This may be
used to check our calculation. So, let us now consider the case of
neutral pions. Isotopic spin invariance forbids the
$\pi^{0}$ and $a_1^{0}$ mesons in the $s$-channel so the lowest allowed state
is the $\omega$ meson.
The $\omega\rho\pi$ vertex has the the form
\beq
i\langle\pi(p)\rho(q)|\omega(p+q)\rangle=
g_{\omega\rho\pi}\epsilon^{\alpha\beta\sigma\tau}
\;\po_{\alpha}^{\omega}\po_{\beta}^{\rho}p_{\sigma}q_{\tau}\;,
\label{orp}
\eeq
where $\po_{\alpha}^{\omega}$ and $\po_{\beta}^{\rho}$ are the polarization
vectors of the $\omega$ and $\rho$ mesons. Then the  contribution of the
$\omega$  to the r.h.s. of eq. (\ref{disp}) is
\beq
\frac{2m_{\rho}^4 Q^2}{(Q^2+m_{\rho}^2)^2 (Q^2+m_{\omega}^2)}
\left( \frac{g_{\omega\rho\pi}}{g_{\rho}}\right) ^{\!2}\;.
\label{om}
\eeq
To be consistent we should use the value of the coupling constant
$g_{\omega\rho\pi}$ obtained from the decay
$\omega\to\pi\gamma$ using $\rho$-dominance \cite{eik},
$g_{\omega\rho\pi}\simeq14.9$ GeV$^{-1}$.
The corresponding $Q^2$ dependence of eq. (24) (multiplied by $Q^4$) is shown
in fig. 2 by the dashed curve for the neutral case.

There is, however, an excited state, $\omega^*(1390)$, which can
contribute. The dominant decay mode is to the $\rho\pi$ channel and taking
this to account for the full width of $230\pm40$ MeV \cite{pdg}, we deduce a
coupling constant $g_{\omega^*\rho\pi}= 5.29$ GeV$^{-1}$.
The result of including both the $\omega$ and $\omega^*$ is shown by the
full line in fig. 2. There is a noticable curvature and a linear fit in this
case results in larger uncertainties: $b=1~-~1.2$ and $c=0~-~0.3$ GeV$^2$.
While the value for $b$ agrees with the one obtained from charged pions,
it is clear that the intercept $c$ is different. This
should have been expected since
$c$ involves the contributions of quark operators and their averages over
charged and neutral pions need not be the same.

Thus, we adopt the value
$b\simeq1.14$, corresponding to a normalization point $\mu\sim Q\sim 1$ GeV.
This is in good agreement with the value $b=1.03$ deduced from
the analysis of ref. \cite{hkl} in which the matrix element
$\langle\pi|\theta_{\mu\nu}^{u+d}|\pi\rangle$ was extracted from a
fit \cite {grv} to the quark and gluon distribution functions in the pion.
In the leading $\log$ approximation the dependence on the normalization point
is determined by the renormalization
group. However, as is well known \cite{pol}, operators of the same twist get
mixed under renormalization due to radiative gluon corrections. The diagonal
combinations in the case of two quark flavors are
\begin{eqnarray}
\theta_{\mu\nu}^{\rm tot}&\pj=
\theta_{\mu\nu}^{u}+\theta_{\mu\nu}^{d}+\theta_{\mu\nu}^{G}~~~~~~&[0]\;,
\nonumber\\
R_{\mu\nu}&\pj=
\theta_{\mu\nu}^{u}+\theta_{\mu\nu}^{d}-\frac{3}{8} \theta_{\mu\nu}^{G}
{}~~~~~~&\left[-{\textstyle\frac{44}{87}}\right]\;,\nonumber\\
\Delta_{\mu\nu}&\hspace{-1.21cm}=\theta_{\mu\nu}^{u}-\theta_{\mu\nu}^{d}~~~~~~
&\left[-{\textstyle\frac{32}{87}}\right]\;.
\label{D}
\end{eqnarray}
The numbers in parentheses are the anomalous dimensions $\gamma$ of the
corresponding diagonal operators which are renormalized multiplicatively,
\beq
\hat O_{Q}=\kappa^{\gamma}\hat O_{\mu}\quad,\quad\kappa=
\frac{\alpha_{s}(\mu^2)}{\alpha_{s}(Q^2)}=
\frac{\log (Q/\Lambda_{QCD})}{\log (\mu/\Lambda_{QCD})}\;,
\label{diag}
\eeq
where $\Lambda_{QCD}\approx 150$ MeV. Then the evolution of $b$, defined by
eq. (\ref{defb}), under a change of the normalization point is given by
\beq
b(\mu)=\frac{16}{11}\left(1-\kappa^{44/87}\right)+b(Q)\kappa^{44/87}\;.
\label{bm}
\eeq
It can be seen that according to eq. (\ref{bm}) $b$ decreases with
$\mu$ and becomes zero at $\mu=1.1\Lambda_{QCD}$. At the
standard normalization point used in QCD sum rules, $\mu=0.5$ GeV, we get
$b=1.06$. Note that the small value of the normalization point for which
$b=0$ (meaning that there is no gluon component in the pion) agrees with the
results of ref. \cite{nsvz} where it was shown that a quark model
description of deep inelastic scattering of leptons on nucleons is consistent
with experimental data provided $\mu\approx m_{\pi}$.

Coming back to finite temperatures, the temperature dependence of
the condensate $\langle{\bf E}^2 +{\bf B}^2\rangle$ is determined by the
integral over the thermal pion phase space
\beq
\langle{\bf E}^2 +{\bf B}^2\rangle_{T}=
3b\int \frac{d^3 p}{(2\pi)^3}\frac{|{\bf p}|}{\exp (|{\bf p}|/T)-1}=
\frac{b\pi^2}{10}T^4\;,
\label{ebt}
\eeq
where the factor of 3 in front of the integral accounts for the three charged
states of pions.
The structure function $C_2$ in eq. (\ref{def}) is obtained in the same way
\beq
C_2(Q,T)=\frac{\pi^2 T^4}{10Q^2}\left(2-b+\frac{\bar{c}}{Q^2}+\ldots\right)+
{\cal O}\left(\frac{T^6}{Q^4}\right)\;.
\label{c2}
\eeq
where $\bar{c}=\frac{2}{3} c_{charged}+\frac{1}{3} c_{neutral}
\approx \frac{2}{3}$ is the charge averaged value of the constant $c$.

Let us now briefly summarize what is known about behavior of condensates at
low temperatures in the chiral limit.
The temperature dependence of the usual (Lorentz scalar) condensates at low $T$
was considered on the basis of chiral perturbation theory up to three-loop
order \cite{leut}. The low $T$ expansion of the quark condensate begins with
a term of order $T^2/F_{\pi}^2$, because for pions with zero momentum
the matrix element $\langle\pi|\bar{q}q|\pi\rangle$ is non-zero and
proportional to $\langle0|\bar{q}q|0\rangle/F_{\pi}^2$. In the case of
the operator $G_{\mu\nu}^{a}G^{a\mu\nu}$,
which is a chiral singlet, the one-pion matrix elements vanish. The $T$
dependence of the gluon condensate is related through the trace anomaly
to $\langle\theta_{\mu}^{\mu}\rangle_T$. The first non-zero contribution to
this matrix element
appears only at the three-loop level. As a result, the $T$ dependence of the
gluon condensate begins at order $T^8/F_{\pi}^4$,
\beq
\langle\frac{\alpha_{s}}{\pi}G^2_{\mu\nu}\rangle_T=
\langle\frac{\alpha_{s}}{\pi}G^2_{\mu\nu}\rangle_0
-\frac{4\pi^2}{3645}N_{f}^{2}
(N_{f}^{2}-1)\frac{T^8}{F_{\pi}^4}
\left(\log\frac{\Lambda_p}{T}-\frac{1}{4}\right)+\ldots\quad,
\label{g2}
\eeq
where $\Lambda_p\simeq 275$ MeV is a scale encountered
in the three-loop calculation of the pressure of a hot pion gas within
chiral perturbation theory \cite{leut}.
The sign of this contribution corresponds to the melting of the gluon
condenate with rising temperature. However,
this melting is much slower than in the case of the quark condensate, and
$\langle G^2\rangle_T$ is practically constant up to $T\sim 150$ MeV, that is,
in the region of applicability of the approximation of a hadronic gas.

One-pion matrix elements of Lorentz non-scalar operators cannot be estimated
using the soft pion approach, because they are proportional to the pion
momentum
$p$. Since $p\sim T$, the corresponding condensates naturally vanish as
$T\to 0$. Since
$\langle{\bf B}^2-{\bf E}^2\rangle_T\simeq\langle{\bf B}^2-{\bf E}^2\rangle_0$,
we get from
eq. (\ref{ebt}) the $T$ dependence of the condensates of chromomagnetic and
chromoelectric fields,
\beq
\langle{\bf B}^2\rangle_T
=\langle{\bf B}^2\rangle_0 +\frac{b\pi^2}{20}T^4\;, ~~~~
\langle{\bf E}^2\rangle_T
=\langle{\bf E}^2\rangle_0 +\frac{b\pi^2}{20}T^4\;,
\label{be}
\eeq
where $\langle{\bf B}^2\rangle_0=-\langle{\bf E}^2\rangle_0\simeq 2\times
10^{-2}$ GeV$^4$, using a renormalization scale $\mu=0.5$ GeV. We indicate
the predicted ratios
$\langle{\bf B}^2\rangle_T/\langle{\bf B}^2\rangle_0$ and
$\langle{\bf E}^2\rangle_T/\langle{\bf E}^2\rangle_0$ by the dashed curves
in fig. 3. It is seen that the
$T$ dependence is rather weak at low $T$ and, at $T\sim 150$ MeV, the
condensates are changed from their $T=0$ value by only about 1\%.
The fact that the change is small is qualitatively  consistent with the
results extracted from the lattice data \cite{ahz}, however we do not agree
with the lattice predictions for the sign. We suggest that the lattice
calculations are probably not sufficiently accurate to predict such small
effects.

 We notice that keeping  $m_{\pi}$ finite would not affect the values
of $b$ and $c$ within the accuracy of our approach. The only differences would
appear in the integral over the thermal distribution function, eq. (28), and
in a lower order contribution to eq. (30). It is straightforward to perform
the calculation numerically and this results in the full curves shown in
fig. 3. We observe that eq. (31) is a good approximation, indeed for the
electric field the results are indistinguishable.
We remark that at very low $T$,  $T\ll m_{\pi}$, we have for $\mu=0.5$ GeV
\beq
\langle{\bf B}^2\rangle_T
=\langle{\bf B}^2\rangle_0 -0.033m_{\pi}^{5/2}T^{3/2}e^{-m_{\pi}/T}\;, ~~~~
\langle{\bf E}^2\rangle_T
=\langle{\bf E}^2\rangle_0 +0.20m_{\pi}^{5/2}T^{3/2}e^{-m_{\pi}/T}
\label{m3}
\eeq
The numerical effect is exceedingly small, but it is interesting to observe
that the magnetic condensate $\langle{\bf B}^2\rangle_T$ slightly
decreases at very low $T$ before increasing. The
behavior of $\langle{\bf E}^2\rangle_T$ is, however, monotonic.

Finally we briefly comment on the effects of
higher spin and twist operators.
The averages of Lorentz non-singlet operators of spin larger than 2 are
necessarily  proportional to higher powers of $T$ and their contribution to
thermal correlators will be suppressed by powers of $T^2/Q^2$. The operators
of spin 2, but of higher twist, are suppressed by $\mu_{h}^2/Q^2$, where
$\mu_{h}$ is some hadronic mass scale $\sim \Lambda_{QCD}$. In the case of
vector currents three twist 4, spin 2 operators \cite{sv}
contribute to the constant
$c$ in eq. (\ref{disp}) and to disentangle individual contributions some
extra information must be used. In our opinion, this problem deserves
further consideration.

We  acknowledge useful discussions with A. Gorski, M. Shifman,
E. Shuryak, C.S. Song and A. Vainshtein. V.E. would like to thank the staff of
the Nuclear Theory Group and Theoretical Physics Institute,
especially Larry McLerran, for the warm hospitality extended
to him during his stay at the University of Minnesota. This work was
supported in part by the  Department of Energy under contract
DE-FG02-87ER40328.

\pagebreak

{\large \bf Figure Captions:}
\begin{itemize}
\item
Fig.  1:  Diagrams contributing to
         (a) $\langle\pi|\theta_{\mu_1\mu_2}^q|\pi\rangle$ and
         (b) $\langle\pi|\theta_{\mu_1\mu_2}^G|\pi\rangle$. The dashed lines
         correspond to gluons.
\item
Fig.  2: The r.h.s. of eq. (\ref{disp}) multiplied by $Q^4$ shown as a
         function of $Q^2$. In the case of charged pions, the dashed curve
         is obtained with $\pi$- and $a_1$-meson intermediate states and
         the full curve also includes the $\pi^{\ast}$
         meson. In the case of neutral pions, the dashed curve
         is obtained with an $\omega$-meson intermediate state and
         the full curve also includes the $\omega^*$ meson.
\item
Fig.  3: The curves labelled B and E give, repectively,
$\langle{\bf B}^2\rangle_T/\langle{\bf B}^2\rangle_0$ and
$\langle{\bf E}^2\rangle_T/\langle{\bf E}^2\rangle_0$ as a function of
temperature. Normalization point is $\mu=0.5$ GeV. The dashed curves give
the results for zero pion mass and the
full curves correspond to non-zero pion mass. In case E these two curves
are indistinguishable.
\end{itemize}
\end{document}